# Changes in Liver Fibrosis in Patients with Chronic Hepatitis B Treated with Pegylated Interferon Combined with Oral Antiviral Agents: A 48-Week Observation from a Prospective Cohort Study

Jinhua Zhao[1,2 #], Lili Wu[1,2], Xiaoan Yang[1,2], Xiaoquan Liu[1,2]，Zhiping Wan[1,2], Bingliang Lin[1,2], Zhiliang Gao[1,2 *], Hong Deng[1,2 *]

[1]Department of Infectious Diseases, Third Affiliated Hospital of Sun Yat-sen University, Guangzhou, China

[2]Guangdong Provincial Key Laboratory of Liver Disease Research, Third Affiliated Hospital of Sun Yat-sen University, Guangzhou, China

*Corresponding authors
Hong Deng, Department of Infectious Diseases, Third Affiliated Hospital of Sun Yat-sen University, 510630, Guangzhou, China, Tel: +86-2085252506, Fax: +86-2085252063.

Email: dhong@mail.sysu.edu.cn

Zhiliang Gao, Department of Infectious Diseases, The Third Affiliated Hospital of Sun Yat-sen University, No 600 Tianhe Road, Guangzhou, 510630, Guangdong Province, China; Telephone: +8620-85252370, Fax: +8620-85252250.

Email: gaozhl@mail.sysu.edu.cn

Funding information
National Natural Science Foundation of China, grant numbers 81870597 and 82170612;
Guangzhou Science and Technology Program Key Projects, grant number: 2023B01J1007.

Authors' contributions

Z.Jh. and W.Ll. proposed the study. Z.Jh.　W.Ll. and D.H designed the study. Z.Jh. and W.Ll.performed research and wrote the first draft manuscript. Z.Jh. W.Ll. and Y.Xa. W.Zp. collected and analyzed the data. D.H. G.Zl.and L.Bl.reviewed the draft manuscript and provided feedback. All authors read and approved the final manuscript.

Acknowledgements

We would like to thank the participants and staff of the Third Affiliated Hospital of Sun Yat-sen University for their valuable contributions.

Funding information
This work was supported by the National Natural Science Foundation of China, grant numbers 81870597 and 82170612, and the Guangzhou Science and Technology Program Key Projects, grant number 2023B01J1007. The funders had no role in the design or conduct of the study; collection, management, analysis or interpretation of the data; or preparation, review or approval of the manuscript.

Ethics declarations
Due to the retrospective nature of the study, the need to obtain the informed consent was waived by the Full name of the IRB/ethical committee.The study protocol was consistent with the International Conference on Harmonization Guidelines, applicable regulations, and ethical guidelines of the Declaration of Helsinki. The protocol and consent forms were approved by the Research Ethics Committee of the Third Affiliated Hospital of Sun Yat-sen University, China ([2018]02-218-08). Patients receiving antiviral therapy were obtained from a real-world study registered at clinicaltrials.gov (chictr180020369).

Data availablity
The datasets analyzed during the current study are not publicly available due privacy reasons but are available from the author zhaojh36@mail2.sysu.edu.cn on reasonable request.

Competing interests
The authors declare that they have no competing interests.

Consent to Participate:Surely.
Consent for Publication:Surely.

Abstract

Background and Aims:
Pegylated interferon (PEG-IFN) combined with oral antiviral agents is currently the most widely used and highly effective treatment regimen for chronic hepatitis B virus (HBV) infection. While effectively suppressing HBV replication, its impact on liver histopathological fibrosis and inflammation remains a critical concern for clinicians and patients. This real-world study evaluated changes in liver fibrosis and inflammation levels before and after 48 weeks of treatment, along with assessments of renal function impairment during therapy.

Methods :
A total of 625 patients who completed 48 weeks of PEG-IFN combined with oral antiviral therapy were enrolled in this real-world study. Based on their virological response at 48 weeks, patients were categorized into Clearance group and Non-clearance group. Changes in liver biochemistry, fibrosis, and renal function were compared between groups and before/after treatment.

Results:
No significant differences were observed in baseline blood tests, liver biochemical markers, or histopathological features between the Clearance group and Non-clearance group. Similarly, baseline renal function showed no significant variation. Further analysis revealed that the Clearance group exhibited significant aggravation of liver fibrosis after 48 weeks of treatment, which correlated strongly with alterations in liver enzyme levels. However, one patient who underwent paired liver biopsies before and after treatment demonstrated marked histopathological improvement in fibrosis. This finding underscores the irreplaceable role of liver histopathology in assessing fibrosis and inflammation. No significant impact on renal function was observed after 48 weeks of treatment.

Conclusion:
PEG-IFN combined with oral antiviral therapy exerts favorable effects on liver fibrosis and inflammation in chronic HBV patients. Non-invasive fibrosis assessment models can monitor fibrotic progression but are susceptible to confounding by hepatic inflammation. Liver histopathology remains the gold standard for evaluation. Long-term follow-up is warranted to assess sustained treatment effects.

Chronic hepatitis B (CHB) and its progression to cirrhosis represent a profound global public health concern, posing significant threats to patients' quality of life and life expectancy [1]. Persistent hepatitis B virus (HBV) infection leads to hepatic inflammation and fibrosis, ultimately progressing to cirrhosis, a condition characterized by gradual loss of liver function and increased risk of complications [2]. Until recently, hepatic fibrosis and cirrhosis were generally considered irreversible [3,4]. However, emerging evidence from serial liver biopsy (LB) studies suggests that certain therapeutic interventions can induce histological improvement, including fibrosis resolution [5-9]. Among these interventions, nucleos(t)ide analogs (NAs), which suppress HBV replication, achieve HBV DNA clearance in 30%–70% of cases and are significantly associated with reduced hepatic inflammation, cellular damage, and fibrosis regression [11-14]. In this study, patients were subjected to combination therapy with pegylated interferon (PEG-IFN) and NAs. While this regimen demonstrates high efficacy in viral clearance, its impact on liver fibrosis remains underexplored. This real-world prospective cohort study evaluates the effects of 48-week antiviral therapy on liver fibrosis.

Given that the prognosis and treatment of CHB and other chronic liver diseases (CLDs) largely depend on the degree of fibrosis, assessing fibrosis regression is clinically valuable [15]. Although liver biopsy remains the gold standard for fibrosis diagnosis, its applicability is limited due to invasiveness, anesthesia-related complications, bleeding risks, and sampling errors. Noninvasive evaluation methods, including the aspartate aminotransferase-to-platelet ratio index (APRI), fibrosis-4 (FIB-4) score, and liver stiffness measurement (LSM), are widely recommended in clinical practice and guidelines [16,17].

PEG-IFN, a broad-spectrum antiviral agent, activates immune responses to suppress HBV replication and enhance viral clearance. Studies indicate that PEG-IFN combined with NAs improves clinical outcomes in CHB patients, including liver function, viral load reduction, and partial attenuation of fibrosis [18]. The synergistic effects of PEG-IFN and oral antiviral agents on liver fibrosis warrant further investigation.

Patients and Methods
Patients
This prospective cohort study enrolled 625 patients, stratified into Clearance group (n=347) and Non-clearance group (n=278) groups based on treatment response at 48 weeks. Baseline fibrosis levels were mild overall; 53 patients underwent initial liver histopathology, with one patient receiving repeat histopathology at 48 weeks, demonstrating clinically significant improvement. All patients met the inclusion criteria of the Everest Project:
1. Clinically diagnosed CHB. 2. Age 18-60 years. 3. ≥1 year of NA therapy, with current HBsAg ≤1500 IU/mL, HBeAg negativity, and HBV DNA <100 IU/mL. 4. No contraindications to interferon, willingness to receive PEG-IFN, and signed informed consent. Exclusion criteria: 1. Interferon allergy. 2. ALT >10× upper limit of normal (ULN) or total bilirubin >2×ULN. 3. Decompensated cirrhosis or history thereof. 4. Leukopenia and/or thrombocytopenia below normal thresholds. 5. Severe cardiovascular, pulmonary, renal, cerebral, or ocular comorbidities. 6. Concurrent autoimmune diseases, psychiatric disorders, diabetes, or thyroid dysfunction. 7. Confirmed/suspected hepatocellular carcinoma or other malignancies. 8. Post-organ transplant or planned transplantation. 9. Immunosuppressant use. 10. Pregnancy or planned pregnancy within 2 years. 11. Alcohol or substance abuse. 12. HIV coinfection. 13. Other contraindications per investigator judgment. Demographic and clinical data included family/medical history of HBV, age, sex, BMI, liver histopathology, and LSM. Laboratory parameters comprised HBsAg, white blood cells, hemoglobin, platelets, neutrophils, lymphocytes, AST, ALT, TBIL, DBIL, GGT, ALB, and GLB. Follow-up assessments were conducted every 12 weeks, with comparative analyses of baseline and 48-week data. Both groups received PEG-IFN combined with oral NAs. Cure status was defined by virological and biochemical responses at 48 weeks.

Liver Stiffness Measurement
The propagation velocity of elastic waves in the liver is directly correlated with hepatic tissue stiffness. Greater tissue stiffness results in faster elastic wave propagation, leading to higher elasticity values (expressed in kilopascals, kPa) displayed by the device, thereby enabling the assessment of liver fibrosis. Ten valid measurements were performed on each patient. The success rate was calculated as the number of valid measurements divided by the total number of measurements. The results were expressed in kilopascals (kPa). The interquartile range (IQR) was defined as an index of intrinsic variability of liver stiffness measurement (LSM), corresponding to the interval around the LSM result encompassing 50% of the valid measurements between the 25th and 75th percentiles. The median value was considered representative of the liver's elastic modulus. Only procedures with ten valid measurements, a success rate of at least 60%, and an IQR-to-median ratio below 30% were deemed reliable. Interpretation of Liver Stiffness Values:<7.3 kPa: Normal, no liver fibrosis (pathological stage S0) ,7.3-9.7 kPa: Pathological stage S1,9.7-12.4 kPa: Pathological stage S2,12.4-17.5 kPa: Pathological stage S3,>17.5 kPa: Pathological stage S4.

Other Non-Invasive Methods for Liver Fibrosis Assessment
Additional non-invasive methods, including the age-platelet index (API), aspartate aminotransferase (AST)-to-platelet ratio index (APRI), and fibrosis-4 index (FIB-4), were evaluated for comparison with LSM results. The API, APRI, and FIB-4 were calculated according to the methods described by Poynard, Wai, and Sterling, respectively (Table 1) [19-21]. Liver and kidney function parameters, including serum alanine aminotransferase (ALT), serum creatinine, and blood urea nitrogen, were measured using a fully automated biochemical analyzer. In accordance with the manufacturer's instructions, the upper limit of normal (ULN) for ALT was defined as 40 IU/L.

Table1.Calculation method of noninvasive model

| Models | Calculation method |
|---|---|
| API | Age（years）:＜30=0,30-39=1,40-49=2,50-59=3,60-69=4,≥70=5 |
| | PLT($10^9$/L-1):≥225=0,200-224=1,175-199=2,150-174=3； |
| | 125-149=4，＜125=5 |
| | API is the sum of the above (possible value 0-10) |
| APRI | [(AST/ULN)/PLT ($10^9$/L-1)]×100 |
| FIB-4 | （Age×AST）÷（PLT×√ALT）×100 |

API:age-platelet index, APRI:aspartate aminotransferase-to-platelet ratio index,FIB-4:age-ALT-platele index

Liver Histological Examination

Liver biopsies were performed via the percutaneous echo-assisted technique, with a minimum requirement of 6 portal tracts. The slides were examined and interpreted by two pathologists at the Third Affiliated Hospital of Sun Yat-sen University. Biopsies were categorized into stages based on the Scheuer scoring system [22]: G0-4 and S0-4.

Statistical analysis

Patient characteristics are presented as mean±standard or number(%) deviation.The independent t test and Fisher's exact test or Mann-Whitney U test were used to compare the baseline characteristics of Clearance group and Non-clearance group,The paried sample t test and McNemar test were used to compare the values pre- and post-antivrial treatment.A two-sided P ＜ 0.05 was considered to be statistically significant.All statistically analyses were performed with IBM SPSS Statistics 27, GraphPad Prism 9.

Results

Baseline Characteristics

Table 2 presents the general demographic and histological characteristics of patients in the Clearance group and Non-clearance group at 48 weeks of treatment. At the initiation of antiviral therapy, the mean age, alanine aminotransferase (ALT) level, and liver stiffness measurement (LSM) value of all patients were 41.02 ± 8.82 years, 36.51 ± 32.60 IU/L, and 7.52 ± 2.41 kPa, respectively. The mean hepatitis B surface antigen (HBsAg) level was 469.29 ± 540.58 IU/L. The calculated mean indices for the Fibrosis-4 (FIB-4) score, aspartate aminotransferase-to-platelet ratio index (APRI), and age-platelet index (API) were 0.01 ± 0.01, 0.35 ± 0.53, and 3.21 ± 1.87, respectively. A total of 53 patients underwent liver histopathological examination, with fibrosis and inflammation grades reported separately for both groups. The mean baseline estimated glomerular filtration rate (eGFR) at treatment initiation was 103.31 ± 15.55. No statistically significant differences were observed between the Clearance group and Non-clearance group in any of the analyzed variables.

Table2.Baseline characteristics of the patients.

| Variables | Total (n=625) | Clearance Group (n=347) | Non-clearance Group (n=278) | P Value |
|---|---|---|---|---|
| Age | 41.02±8.82 | 39.15±8.52 | 43.35±8.66 | 0.830 |
| BMI | 23.28±4.15 | 23.32±4.33 | 23.22±3.93 | 0.411 |
| Years since hepatitis B | 17.03±9.09 | 17.24±9.16 | 16.75±9 | 0.977 |
| History of hepatitis B | 242（39%） | 132（38%） | 110(40%) | 0.464 |
| History of Drinking | 23(4%) | 8(2%) | 15(5%) | 0.107 |
| High blood pressure | 14(2%) | 6(2%) | 8(3%) | 0.585 |
| Diabetes | 12(2%) | 4(1%) | 8(3%) | 0.257 |
| HBsAg (IU/L) | 469.29±540.58 | 243.58±283.47 | 751.01±643.79 | ＜0.001 |
| WBC ($10^9$/L) | 5.99±1.53 | 5.88±1.56 | 6.13±1.48 | 0.053 |
| HGB (g/L) | 152.93±17.32 | 152.52±16.91 | 153.43±17.82 | 0.216 |
| PLT ($10^9$/L) | 207.89±56.26 | 208.80±58.96 | 206.74±52.77 | 0.090 |
| NEUT ($10^9$/L) | 3.36±1.22 | 3.27±1.23 | 3.48±1.19 | 0.310 |
| LYMP ($10^9$/L) | 2.01±0.58 | 1.98±0.58 | 2.04±0.58 | 0.993 |
| AST (U/L) | 31.27±22.96 | 31.07±25.92 | 31.52±18.67 | 0.899 |
| ALT (U/L) | 36.51±32.60 | 36.80±37.17 | 36.15±25.84 | 0.556 |
| TBIL （μmol/L） | 11.90±5.85 | 11.65±5.04 | 12.21±6.73 | 0.045 |
| DBIL （μmol/L） | 3.42±2.80 | 3.24±1.66 | 3.65±3.76 | 0.003 |
| GGT （U/L） | 33.58±25.96 | 31.88±23.50 | 35.71±28.64 | 0.706 |
| LDH | 190.83±35.02 | 187.77±34.33 | 194.66±35.56 | 0.889 |
| ALB （g/L） | 47.93±2.96 | 48.20±2.74 | 47.58±3.18 | 0.023 |
| GLB （g/L） | 27.31±4.15 | 27.59±4.36 | 26.96±3.85 | 0.188 |
| LSM （kPa） | 7.52±2.41 | 7.59±2.49 | 7.44±2.3 | 0.097 |
| FIB-4 | 0.01±0.01 | 0.01±0.01 | 0.01±0.01 | 0.386 |
| APRI | 0.35±0.53 | 0.36±0.69 | 0.32±0.19 | 0.062 |
| API | 3.21±1.87 | 3.03±1.88 | 3.43±1.84 | 0.699 |
| Initial biopsy | 53（8.5%） | 36（10.4%） | 17（6.1%） | |
| Grade | | | | 0.154 |
| G0 | 2（3.8%） | 2（5.6%） | 0（0） | |
| G1 | 34（64.2%） | 25（69.4%） | 9（52.9%） | |
| G2 | 13（24.5%） | 8（22.2%） | 5（29.4%） | |
| G3 | 2（4.8%） | 0（0） | 2（11.8%） | |
| Stage | | | | 0.154 |
| S0 | 14（26.4%） | 9（25.0%） | 5（29.4%） | |
| S1 | 18（34.0%） | 14（38.9%） | 4（23.5%） | |
| S2 | 11（20.8%） | 7（19.4%） | 4（23.5%） | |
| S3 | 7（13.2%） | 5（13.9%） | 2（11.8%） | |
| S4 | 2（3.8%） | 1（2.8%） | 1（5.9%） | |
| eGFR (ml/min/1.73m$^2$) | 103.31±15.55 | 104.64±16.10 | 101.64±14.70 | 0.221 |
| Urea (mmol/L) | 4.88±1.10 | 4.77±1.09 | 5.02±1.09 | 0.492 |
| CREAT（μmol/L） | 76.83±15.30 | 76.38±15.84 | 77.40±14.62 | 0.132 |
| Cysc （mg/L） | 0.88±0.14 | 0.87±0.14 | 0.90±0.13 | 0.051 |

**Comparison of Pre- and Post-Antiviral Treatment Values in the Clearance Group**

Table 3 summarizes the laboratory test results, liver stiffness measurement (LSM) values, and several non-invasive models for assessing liver fibrosis in the Clearance group before and after antiviral therapy. Compared to baseline, the patients exhibited varying degrees of elevation in AST and ALT levels (P=0.036 and P=0.461, respectively). Among the non-invasive liver fibrosis assessment indices, LSM, API, APRI, and FIB-4 all showed increases, with statistically significant differences. The changes in these fibrosis-related indices are illustrated in Figure 1.

Table3.Comparison between pre- and post-antiviral treatment.

| | Clearance Group（n=347） | | |
|---|---|---|---|
| | Pre | Post | P value |
| HBsAg (IU/L) | 243.58±283.47 | 0.88±0.45 | ＜0.001 |
| AST (U/L) | 31.07±25.92 | 36.00±35.22 | 0.036 |
| ALT (U/L) | 36.80±37.17 | 39.04±42.41 | 0.461 |
| TBIL（μmol/L） | 11.65±5.04 | 10.10±3.89 | ＜0.001 |
| WBC (10$^9$/L) | 5.88±1.56 | 4.12±1.51 | ＜0.001 |
| PLT (10$^9$/L) | 208.80±58.96 | 156.53±60.85 | ＜0.001 |
| LYMP (10$^9$/L) | 1.98±0.58 | 1.46±0.59 | ＜0.001 |
| NEUT (10$^9$/L) | 3.27±1.23 | 2.17±1.09 | ＜0.001 |
| LSM（kPa） | 7.59±2.49 | 8.21±2.10 | ＜0.001 |
| FIB-4 | 0.01±0.014 | 0.02±0.03 | ＜0.001 |
| APRI | 0.36±0.69 | 0.62±1.55 | 0.005 |
| API | 3.03±1.88 | 4.64±1.98 | ＜0.001 |

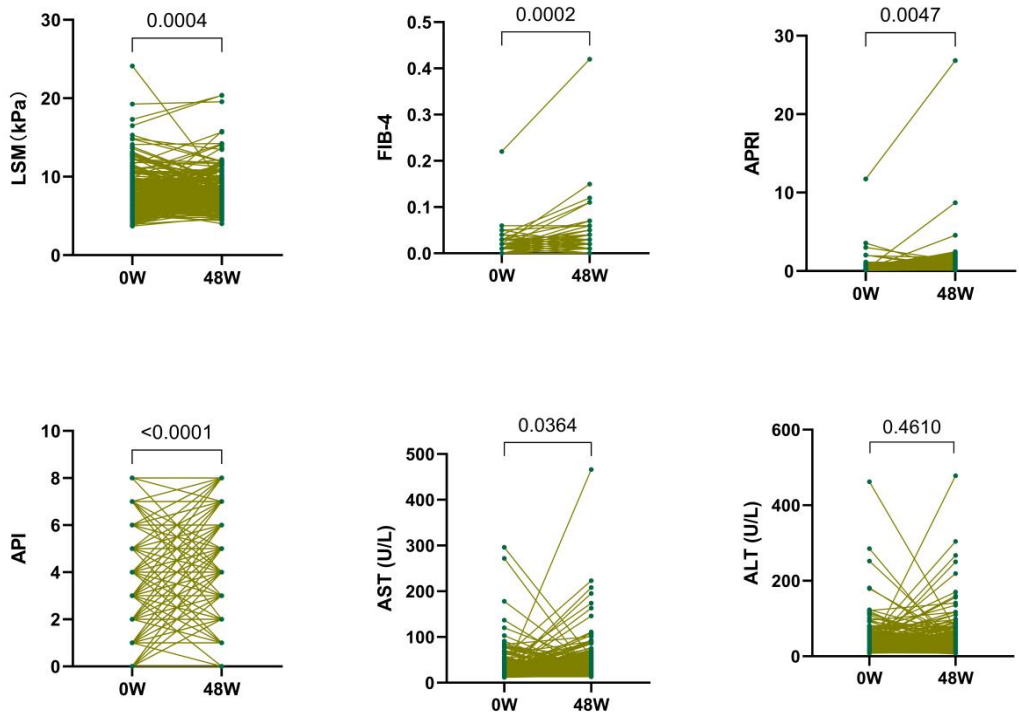

Figure1.Variation in LSM、FIB-4、APRI and API values and AST、ALT levels between pre- and post-antiviral treatment in patients with CHB who underwent follow-up LSM after 1 year of antiviral treatment using PEG-IFN with NAs.

## Factors Influencing Changes in Liver Stiffness Measurement (LSM) at Week 48 in Patients with Chronic Hepatitis B

Based on the changes in LSM at week 48, the patients were categorized into two groups: a decreased LSM group and an increased LSM group. General demographic data, liver function biochemical parameters, and metabolic-related indicators were included in univariate and multivariate binary logistic regression models. The results indicated that the alanine aminotransferase (ALT) level at week 36 was a significant influencing factor for LSM outcomes (OR = 1.006, P = 0.042), as shown in Table 4.

Table 4. Factors Influencing Changes in LSM

| Characteristics | Total(N) | Univariate analysis | | Multivariate analysis | |
|---|---|---|---|---|---|
| | | Odds Ratio (95% CI) | P value | OddsRatio(95%CI) | P value |
| Sexl | 614 | | | | |
| 1（Male） | 536 | Reference | | Reference | |
| 2(Female) | 78 | 0.645 (0.394 – 1.056) | 0.081 | 0.677 (0.401 – 1.142) | 0.143 |
| Family History of Hepatitis B | 481 | | | | |
| 1 | 235 | Reference | | | |
| 0 | 246 | 0.994 (0.680 – 1.452) | 0.974 | | |
| Alcohol Consumption History | 482 | | | | |
| 0 | 460 | Reference | | | |
| 1 | 22 | 1.329 (0.510 – 3.464) | 0.561 | | |
| Hypertension | 484 | | | | |
| 0 | 470 | Reference | | | |
| 1 | 14 | 1.254 (0.387 – 4.062) | 0.706 | | |
| Diabetes | 483 | | | | |
| 0 | 471 | Reference | | | |
| 1 | 12 | 1.000 (0.297 – 3.372) | 1.000 | | |
| Age | 614 | 1.025 (1.005 – 1.045) | 0.014 | 1.017 (0.995 – 1.039) | 0.132 |
| LSM-0W | 614 | 1.028 (0.955 – 1.106) | 0.464 | | |
| AST-0W | 614 | 1.007 (0.997 – 1.017) | 0.161 | | |
| AST-12W | 614 | 1.003 (0.998 – 1.007) | 0.256 | | |
| AST-24W | 614 | 0.998 (0.994 – 1.002) | 0.299 | | |
| AST-36W | 614 | 1.001 (0.995 – 1.007) | 0.785 | | |
| AST-48W | 614 | 1.002 (0.997 – 1.008) | 0.370 | | |
| ALT-0W | 614 | 1.006 (0.999 – 1.013) | 0.089 | 1.006 (0.999 – 1.013) | 0.114 |
| ALT-12W | 614 | 1.002 (0.998 – 1.005) | 0.365 | | |

Table 4. Factors Influencing Changes in LSM

| Characteristics | Total(N) | Univariate analysis | | Multivariate analysis | |
|---|---|---|---|---|---|
| | | Odds Ratio (95% CI) | P value | OddsRatio(95%CI) | P value |

| Characteristics | Total(N) | Univariate analysis | | Multivariate analysis | |
| --- | --- | --- | --- | --- | --- |
| | | Odds Ratio (95% CI) | P value | OddsRatio(95%CI) | P value |
| ALT-24W | 614 | 1.002 (0.999 – 1.006) | 0.216 | | |
| ALT-36W | 614 | 1.005 (0.999 – 1.011) | 0.095 | 1.006 (1.000 – 1.013) | 0.042 |
| ALT-48W | 614 | 1.004 (0.998 – 1.009) | 0.168 | | |
| TBIL-0W | 614 | 1.010 (0.988 – 1.033) | 0.377 | | |
| DBIL-0W | 614 | 1.007 (0.998 – 1.016) | 0.107 | | |
| GGT-0W | 614 | 1.002 (1.000 – 1.004) | 0.027 | 1.001 (0.996 – 1.006) | 0.740 |
| CK-0W | 614 | 1.001 (0.999 – 1.003) | 0.435 | | |
| LDH-0W | 614 | 0.998 (0.996 – 1.001) | 0.163 | | |
| AMY-0W | 614 | 0.998 (0.993 – 1.003) | 0.366 | | |
| ALB-0W | 614 | 1.006 (1.000 – 1.012) | 0.059 | 0.997 (0.977 – 1.017) | 0.757 |
| GLB-0W | 614 | 0.985 (0.971 – 1.000) | 0.050 | 1.000 (0.958 – 1.043) | 0.999 |
| eGFR-0W | 614 | 0.991 (0.982 – 0.999) | 0.028 | 0.990 (0.976 – 1.003) | 0.140 |
| Urea-0W | 614 | 0.959 (0.881 – 1.043) | 0.330 | | |
| CREAT-0W | 614 | 0.997 (0.992 – 1.001) | 0.119 | | |
| IPHOS-0W | 614 | 1.001 (0.972 – 1.031) | 0.957 | | |
| Cysc-0W | 340 | 0.694 (0.145 – 3.320) | 0.647 | | |
| TBA-0W | 340 | 1.003 (0.980 – 1.026) | 0.826 | | |
| Glu-0W | 614 | 1.110 (0.822 – 1.498) | 0.495 | | |
| VitD-0W | 614 | 1.013 (1.001 – 1.025) | 0.027 | 1.011 (0.999 – 1.023) | 0.068 |
| FT3-0W | 614 | 1.009 (0.745 – 1.366) | 0.956 | | |
| FT4-0W | 614 | 0.932 (0.836 – 1.039) | 0.202 | | |
| TSH-0W | 614 | 0.996 (0.792 – 1.252) | 0.974 | | |
| CHOL-0W | 614 | 0.856 (0.639 – 1.147) | 0.297 | | |
| TRIG-0W | 614 | 1.043 (0.719 – 1.512) | 0.825 | | |
| HDLC-0W | 614 | 0.645 (0.226 – 1.840) | 0.413 | | |
| LDLC-0W | 614 | 0.969 (0.674 – 1.395) | 0.867 | | |

| Characteristics | Total(N) | Univariate analysis | | Multivariate analysis | |
|---|---|---|---|---|---|
| | | Odds Ratio (95% CI) | P value | OddsRatio(95%CI) | P value |
| UA-0W | 614 | 1.001 (0.997 – 1.005) | 0.598 | | |

## Changes in eGFR and Related Values Before and After Antiviral Treatment in the Clearance Group

The changes in estimated glomerular filtration rate (eGFR) and related values before and after antiviral treatment were evaluated in the Clearance group (Figure 2). No significant alterations were observed in eGFR or lactate dehydrogenase (LDH) levels before and after treatment. However, among the remaining renal function-related indices, urea (Urea), creatinine (CREAT), cystatin C (CysC), and inorganic phosphate (IPHOS) exhibited statistically significant differences.

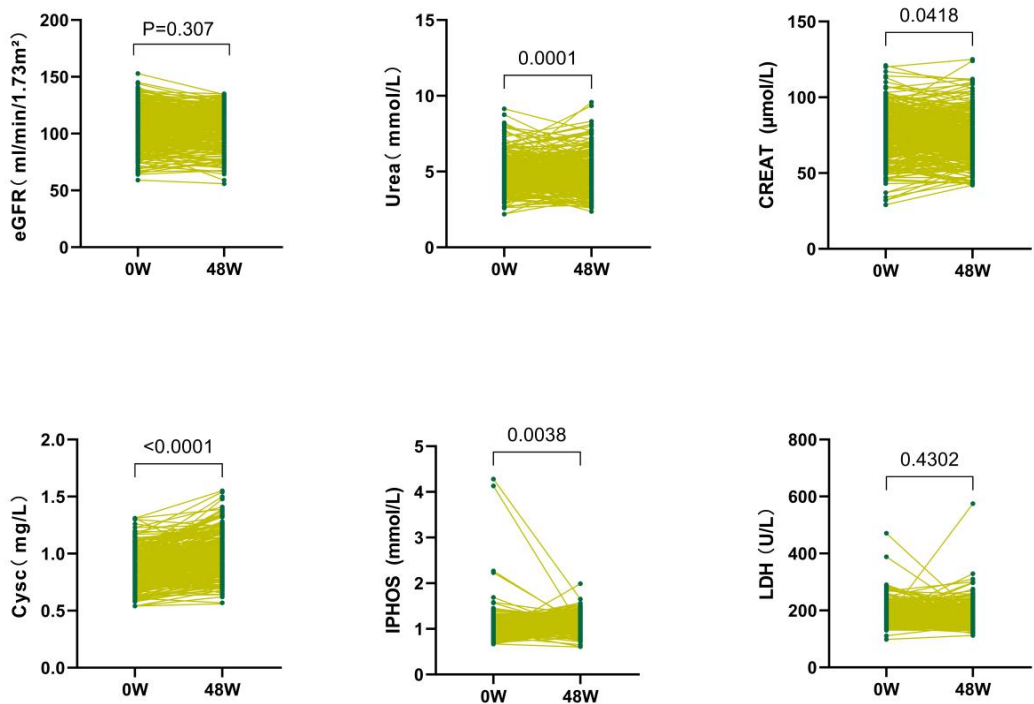

Figure2.Variation in eGFR、Urea、CREAT and Cysc values between pre- and post-antiviral treatment in patients with CHB who underwent follow-up after 1 year of antiviral treatment using PEG-IFN with NAs.

## Discussion

Hepatic fibrosis is a critical pathological process in the progression of chronic hepatitis B virus (HBV) infection, the severity of which not only influences disease prognosis but may also significantly impact the response to antiviral therapy.

Liver biopsy remains the gold standard for assessing the degree and staging of hepatic fibrosis and is frequently performed as a baseline evaluation prior to antiviral therapy [22]. Liver biopsy enables direct observation of fibrosis distribution and architecture while concurrently evaluating other pathological changes. Currently, widely adopted pathological staging systems are based on liver biopsy findings. Additionally, guiding treatment decisions is a primary objective of liver biopsy, as it serves as the only reliable method for diagnosing cirrhosis, which is pivotal in determining treatment plans.Patients with moderate-to-severe fibrosis (F3-F4) may require avoidance of interferon-based therapy.

Several noninvasive hepatic fibrosis models have been developed to predict fibrosis severity [19-21]; however, none have been specifically validated for assessing fibrosis before and after treatment with pegylated interferon (PEG-IFN) combined with oral antiviral agents. Our findings indicate that liver stiffness measurement (LSM), like other noninvasive fibrosis models, demonstrated worsening fibrosis post-treatment, a phenomenon that may be attributed to the robust hepatic inflammatory response during PEG-IFN therapy. Previous studies have reported a significant influence of alanine aminotransferase (ALT) levels on LSM [24-27]. Recent research has also shown that intrahepatic inflammatory activity markedly affects LSM in patients with nonalcoholic fatty liver disease [28]. Since serum fibrosis markers, such as hyaluronic acid and type IV collagen, were not assessed in this retrospective study, future investigations should further explore these biomarkers.

Studies on the impact of antiviral therapy on hepatic fibrosis in chronic hepatitis B (CHB) have demonstrated promising results. A randomized, double-blind, controlled trial revealed that nearly three-quarters of cirrhotic patients achieved cirrhosis regression after five years of tenofovir disoproxil fumarate (TDF) treatment. However, this study had a relatively small sample size, and only a limited number of patients underwent a second liver biopsy [27]. Another study reported that pediatric CHB patients with advanced fibrosis treated with interferon-α for 48 weeks achieved a hepatitis B surface antigen (HBsAg) seroclearance rate of 35.7%, with significant histological improvement observed in all patients [28]. These findings align with studies in adult CHB patients with advanced fibrosis, where long-term nucleos(t)ide analogue (NUC) therapy combined with PEG-IFN led to substantial histological improvement. Moreover, patients who received PEG-IFN add-on therapy exhibited a faster HBsAg decline and higher cumulative HBsAg seroclearance rates.

In our study, enrolled patients had baseline HBsAg levels <1500 IU/mL and were treated with PEG-IFN plus oral antiviral agents. Liver function tests revealed significant alterations in liver enzymes during treatment, predominantly elevated ALT levels, which correlated with increased noninvasive fibrosis scores and post-treatment LSM values. This suggests that hepatic inflammatory activity must be carefully considered when evaluating fibrosis progression in PEG-IFN-treated patients. Notably, one patient who underwent paired liver biopsies before and after treatment exhibited consistent serological changes with other participants but demonstrated remarkable fibrosis regression histologically, reinforcing liver biopsy as the gold standard for hepatic fibrosis assessment.

We acknowledge several limitations in our study. First, the absence of a control group precludes direct comparison of changes in LSM and other parameters from baseline to follow-up. Second, the short follow-up period, coupled with intense hepatic inflammation during the 48-week treatment course, may have confounded noninvasive fibrosis scores and LSM values, limiting their accuracy in reflecting true fibrosis severity. Extending the treatment and follow-up duration would allow for resolution of inflammation and a more reliable fibrosis assessment. Third, patient reluctance to undergo repeat liver biopsies resulted in limited histological follow-up data. Therefore, further studies are warranted to validate the utility of noninvasive fibrosis models and LSM in monitoring fibrosis changes during CHB treatment.

In summary, our preliminary study indicates that the combination of pegylated interferon and oral antiviral agents did not lead to a significant improvement in liver fibrosis among patients with chronic hepatitis B, as initially expected. However, the observation period in this study was relatively short, and antiviral treatment is still ongoing. Therefore, further follow-up is anticipated to observe potential improvements. Non-invasive liver fibrosis models and LSM were well accepted by the majority of patients, while liver histopathological biopsy remains irreplaceable in certain specific clinical scenarios. As a prospective study, we will continue to monitor these patients over the long term to assess changes in liver fibrosis levels following hepatitis B virus clearance. Additionally, we will further refine liver histopathological examinations to validate our conclusions.

Authors' contributions

Z.Jh. and W.Ll. proposed the study. Z.Jh.　W.Ll. and D.H designed the study. Z.Jh. and W.Ll.performed research and wrote the first draft manuscript. Z.Jh. W.Ll. and Y.Xa. W.Zp. collected and analyzed the data. D.H. G.Zl.and L.Bl.reviewed the draft manuscript and provided feedback. All authors read and approved the final manuscript.

Acknowledgements
We would like to thank the participants and staff of the Third Affiliated Hospital of Sun Yat-sen University for their valuable contributions.

Funding information
This work was supported by the National Natural Science Foundation of China, grant numbers 81870597 and 82170612, and the Guangzhou Science and Technology Program Key Projects, grant number 2023B01J1007. The funders had no role in the design or conduct of the study; collection, management, analysis or interpretation of the data; or preparation, review or approval of the manuscript.

Ethics declarations
Due to the retrospective nature of the study, the need to obtain the informed consent was waived by the Full name of the IRB/ethical committee.The study protocol was consistent with the International Conference on Harmonization Guidelines, applicable regulations, and ethical guidelines of the Declaration of Helsinki. The protocol and consent forms were approved by the Research Ethics Committee of the Third Affiliated Hospital of Sun Yat-sen University, China ([2018]02-218-08). Patients receiving antiviral therapy were obtained from a real-world study registered at clinicaltrials.gov (chictr180020369).

Data availablity
The datasets analyzed during the current study are not publicly available due privacy reasons but are available from the author zhaojh36@mail2.sysu.edu.cn on reasonable request.

Competing interests
The authors declare that they have no competing interests.

ORCID
Jinhua Zhao https://orcid.org/0009-0008-1757-8841